# Archaeoastronomy in India

Subhash Kak
Oklahoma State University, Stillwater

Our understanding of archaeoastronomical sites in India is based not only on a rich archaeological record and texts that go back thousands of years, but also on a living tradition that is connected to the past. Conversely, India has much cultural diversity and a tangled history of interactions with neighboring regions that make the story complex. The texts reveal to us the cosmological ideas that lay behind astronomical sites in the historical period and it is generally accepted that the same idea also apply to the Harappan era of the third millennium BCE (Kenoyer, 1998: 52-53).

In the historical period, astronomical observatories were part of temple complexes where the king was consecrated. Such consecration served to confirm the king as foremost devotee of the chosen deity, who was taken to be the embodiment of time and the universe (Kak, 2002a: 58). For example, Udayagiri is an astronomical site connected with the Classical age of the Gupta dynasty (320-500 CE), which is located a few kilometers from Vidisha in central India (Willis, 2001; Dass and Willis, 2002). The imperial Guptas enlarged the site, an ancient hilly observatory going back at least to the 2$^{nd}$ century BCE at which observations were facilitated by the geographical features of the hill, into a sacred landscape to draw royal authority.

Indian astronomy is characterized by the concept of ages of successive larger durations, which is an example of the pervasive idea of recursion, or repetition of patterns across space, scale and time. An example of this is the division of the ecliptic into 27 star segments (*nakshatras*), with which the moon is conjoined in its monthly circuit, each of which is further sub-divided into 27 sub-segments (*upa-nakshatras*), and the successive divisions of the day into smaller measures of 30 units. The idea of recursion underlies the concept of the sacred landscape and it is embodied in Indian art, providing an archaeoastronomical window on sacred and monumental architecture. It appears that this was an old idea because intricate spiral patterns, indicating recursion, are also found in the paintings of the Mesolithic period. Tyagi (1992) has claimed that they are unique to Indian rock art.

According to the *Vāstu Shāstra*, the structure of the building mirrors the emergence of cosmic order out of primordial chaos through the act of measurement. The universe is symbolically mapped into a square that emphasizes the four cardinal directions. It is represented by the square *vāstu-mandala*, which in its various forms is the basic plan for the house and the city. There exist further elaborations of this plan, some of which are rectangular.

It is significant that *yantric* buildings in the form of *mandalas* have been discovered in North Afghanistan that belong to a period that corresponds to the late stage of the Harappan tradition (Kak, 2000a; Kak, 2005b) providing architectural evidence in support of the idea of recursion at this time. Although these building are a part of the Bactria-Margiana Archaeological Complex (BMAC), their affinity with ideas that are also present in the Harappan system shows that these ideas were widely spread.

# Contents

## 1. Chronology and Overview

India's archaeological record in the northwest has unbroken continuity going back to about 7500 BCE at Mehrgarh (Kenoyer, 1998; Lal, 2002), and it has an rock art tradition, next only to that of Australia and Africa in abundance, that is much older (Pandey, 1993; Bednarik, 2000). Some rock art has been assigned to the Upper Paleolithic period. There is surprising uniformity, both in style and content, in the rock art paintings of the Mesolithic period (10,000 – 2500 BCE) (Tyagi, 1992; Wakankar, 1992).

The archaeological phases of the Indus (or Sindhu-Sarasvati) tradition have been divided into four eras: *early food-producing era* (c. 6500- 5000 BCE), *regionalization era* (5000 – 2600 BCE), *integration era* (2600 – 1900 BCE), and *localization era* (1900 – 1300 BCE) (Shaffer, 1992). The early food-producing era lacked elaborate ceramic technology. The regionalization era was characterized by styles in ceramics, lapidary arts, glazed faience and seal making that varied across regions. In the integration era, there is significant homogeneity in material culture over a large geographical area and the use of the so-called Indus script, which is not yet deciphered. In the localization era, patterns of the integration era are blended with regional ceramic styles, indicating decentralization and restructuring of the interaction networks. The localization era of the Sindhu-Sarasvati tradition is the regionalization era of the Ganga-Yamuna tradition which transforms into the integration era of the Magadha and the Mauryan dynasties. There is also continuity in the system of weights and lengths between the Harappan period and the later historic period (Mainkar, 1984).

The cultural mosaic in the third millennium BCE is characterized by the integration phase of the Harappan civilization of northwest India, copper and copper/bronze age cultures or central and north India, and Neolithic cultures of south and east India (Lal, 1997). Five large cities of the integration phase are Mohenjo-Daro, Harappa, Ganweriwala, Rakhigarhi, and Dholavira. Other important sites of this period are Kalibangan, Rehman Dheri, Naushara, Kot Diji, and Lothal.

A majority of the towns and settlements of the Harappan period were in the Sarasvati valley region. Hydrological changes, extended period of drought, and the drying up of the Sarasvati River due to its major tributaries being captured by the Sindh and Ganga Rivers after an earthquake in 1900 BCE led to the abandonment of large areas of the Sarasvati valley (Kak, 1992). The Harappan phase went through various stages of decline during the second millennium BCE. A second urbanization began in the Ganga and Yamuna valleys around 900 BCE. The earliest surviving records of this culture are in Brahmi script. This second urbanization is generally seen at the end of the Painted Gray Ware (PGW) phase (1200- 800 BCE) and with the use of the Northern Black Polished Ware (NBP) pottery. Late Harappan was partially contemporary with the PGW phase. In other words, a continuous series of cultural developments link the two early urbanizations of India.

The setting for the hymns of the *Rigveda*, which is India's most ancient literary text, is the area of Sapta Saindhava, the region of north India bounded by the Sindh and the Ganga rivers although regions around this heartland are also mentioned. The *Rigveda* describes the Sarasvati River to be the greatest of the rivers and going from the mountains to the sea. The archaeological record, suggesting that this river had turned dry by1900 BCE, indicates that the *Rigveda* is prior to this epoch. The *Rigveda* and other early Vedic literature have astronomical references related to the shifting astronomical frame that indicate epochs of the fourth and third millennium BCE which is consistent with the hydrological evidence. The nakshatra lists are found in the Vedas, either directly or listed under their presiding deities, and it one may conclude that their names have not changed. Vedic astronomy used a luni-solar year in which an intercalary month was employed as adjustment with solar year.

The shifting of seasons through the year and the shifting of the northern axis allow us to date several statements in the Vedic books (Sastry, 1985). Thus the *Shatapatha Brāhmana* (2.1.2.3) has a statement that points to an earlier epoch where it is stated that the Krittikā (Pleiades) never swerve from the east. This corresponds to 2950 BCE. The *Maitrāyanīya Brāhmana Upanishad* (6.14) refers to the winter solstice being at the mid-point of the Shravishthā segment and the summer solstice at the beginning of Maghā. This indicates 1660 BCE. The *Vedānga Jyotisha* mentions that winter solstice was at the beginning of Shravishthā and the summer solstice at the mid-point of Ashleshā. This corresponds to about 1300 BCE.

The nakshatras in the Vedānga Jyotisha are defined to be 27 equal parts of the ecliptic. The nakshatra list of the late Vedic period begin with Krittikā (Pleiades) whereas that of the astronomy texts after 200 CE begin with Ashvini ($\alpha$ and $\beta$ Arietis), indicating a transition through 2 nakshatras, or a time span of about 2,000 years.

The foundation of Vedic cosmology is the notions of *bandhu* (homologies or binding between the outer and the inner). In the Ayurveda, medical system associated with the Vedas, the 360 days of the year were taken to be mapped to the 360 bones of the developing fetus, which later fuse into the 206 bones of the person. It was estimated correctly that the sun and the moon were approximately 108 times their respective diameters from the earth (perhaps from the discovery that the angular size of a pole removed 108 times its height is the same as that of the sun and the moon), and this number was used in sacred architecture. The distance to the sanctum sanctorum of the temple from the gate and the perimeter of the temple were taken to be 54 and 180 units, which are one-half each of 108 and 360 (Kak, 2005a). Homologies at many levels are at the basis of the idea of *recursion,* or repetition in scale and time. The astronomical basis of the Vedic ritual was the reconciliation of the lunar and solar years (Kak, 2000a; Kak, 2000b).

Texts of the Vedic and succeeding periods provide us crucial understanding of the astronomy and the archaeoastronomy of the historical period throughout India. The medieval period was characterized by pilgrimage centers that created sacred space mirroring conceptions of the cosmos. Sacred temple architecture served religious and political ends.

The instruments that were used in Indian astronomy include the water clock (*ghati yantra)*, gnomon (*shanku*), cross-staff (*yasti yantra*), armillary sphere (*gola-yantra*), board for sun's altitude (*phalaka yantra*), sundial (*kapāla yantra*), and astrolabe (Gangooly, 1880). In early 18th century, Maharaja Sawai Jai Singh II of Jaipur (r. 1699-1743) built five masonry observatories called Jantar Mantar in Delhi, Jaipur, Ujjain, Mathura, and Varanasi. The Jantar Mantar consists of the Ram Yantra (a cylindrical structure with an open top and a pillar in its center to measure the altitude of the sun), the Rashivalaya Yantra (a group of twelve instruments to determine celestial latitude and longitude), the Jai Prakash (a concave hemisphere), the Laghu Samrat Yantra (small sundial), the Samrat Yantra (a huge equinoctial dial), the Chakra Yantra (upright metal circles to find the right ascension and declination of a planet), the Digamsha Yantra (a pillar surrounded by two circular walls), the Kapali Yantra (two sunken hemispheres to determine the position of the sun relative to the planets and the zodiac), and the Narivalaya Yantra (a cylindrical dial).

## 2. Pre-historical and Harappan Period

The city of Mohenjo-Daro (2500 BCE), like most other Harappan cities (with the exception of Dholavira as far as we know at this time) was divided into two parts: the acropolis and the lower city. The Mohenjo-Daro acropolis, a cultural and administrative centre, had as its foundation a 12 meter high platform of 400 m × 200 m. The lower city had streets oriented according to the cardinal directions and provided with a network of covered drains. Its houses had bathrooms. The city's wells were so well constructed with tapering bricks that they have not collapsed in 5000 years. The Great Bath (12 m × 7 m) was built using finely fitted bricks laid on with gypsum plaster and made watertight with bitumen. A high corbelled outlet allowed it to be emptied easily. Massive walls protected the city against flood water.

The absence of monumental buildings such as palaces and temples makes the Harappan city strikingly different from its counterparts of Mesopotamia and Egypt, suggesting that the polity of the Harappan state was de-centralized and based on a balance between the political, the mercantile, and the religious elites. The presence of civic amenities such as wells and drains attests to considerable social equality. The power of the mercantile guilds is clear in the standardization of weights of carefully cut and polished chart cubes that form a combined binary and decimal system.

Mohenjo-Daro and other sites show slight divergence of 1° to 2° clockwise of the axes from the cardinal directions (Wanzke, 1984). It is thought that this might have been due to the orientation of Aldebaran (*Rohinī* in Sanskrit) and the Pleiades (*Krtikkā* in Sanskrit) that rose in the east during 3000 BCE to 2000 BCE at the spring equinox; the word "rohinī" literally means rising. Furthermore, the slight difference in the orientations amongst the buildings in Mohenjo-Daro indicates different construction periods using the same traditional sighting points that had shifted in this interval (Kenoyer, 1998).

Mohenjo-Daro's astronomy used both the motions of the moon and the sun (Maula, 1984). This is attested by the use of great calendar stones, in the shape of ring, which served to mark the beginning and end of the solar year.

*Dholavira*
Dholavira is located on an island just north of the large island of Kutch in Gujarat. Its strategic importance lay in its control of shipping between Gujarat and the delta of the Sindh and Sarasvati rivers.

The layout of Dholavira is unique in that it comprises of three "towns," which is in accord with Vedic ideas (Bisht, 1997; Bisht, 1999a; Bisht, 1999b). The feature of recursion in the three towns, or repeating ratios at different scales, is significant. Specifically, the design is characterized by the nesting proportion of 9:4 across the lower and the middle towns and the castle. The proportions of 5/4, 7/6, and 5/4 for the lower town, the middle town, and the castle may reflect the measures related to the royal city, the commander's quarter, and the king's quarter, respectively, which was also true of Classical India (Bhat, 1995).

A Dholavira length, D, has been determined by finding the largest measure which leads to integer dimensions for the various parts of the city. This measure turns out be the same as the *Arthaśāstra* (300 BCE) measure of *dhanus* (arrow) that equals 108 *angulas* (fingers). This scale is confirmed by a terracotta scale from Kalibangan and the ivory scale found in Lothal. The Kalibangan scale (Joshi, 2007; Balasubramaniam and Joshi, 2008) corresponds to units of 17.5 cm, which is substantially the same as the Lothal scale and the small discrepancy may be a consequence of shrinkage upon firing.

The analysis of the unit of length at Dholavira is in accord with the unit from the historical period (Danino, 2005; Danino, 2008). The unit that best fits the Dholavira dimensions is 190.4 cm, which when divided by 108 gives the Dholavira *angula* of 1.763 cm. The subunit of *angula* is confirmed when one considers that the bricks in Harappa follow ratios of 1:2:4 with the dominating size being 7 × 14 × 28 cm (Kenoyer, 1998). These dimensions can be elegantly expressed as 4 × 8 × 16 *angulas*, with the unit of *angula* taken as 1.763 cm. It is significant that the ivory scale at Lothal has 27 graduations in 46 mm, or each graduation is 1.76 mm.

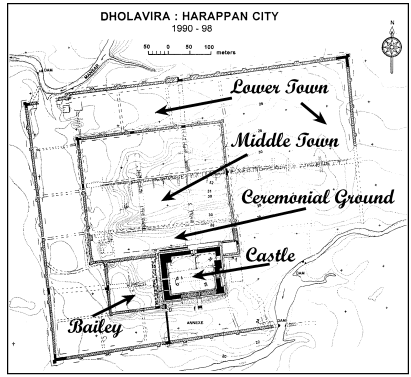

Figure 1. Map of Dholavira (Bisht, 1997)

With the new Dhloavira unit of D, the dimensions of Mohenjo-Daro's acropolis turn out to be 210 × 105 D; Kalibangan's acropolis turn out to be 126 × 63 D. The dimensions of the lower town of Dholavira are 405 × 324 D; the width of the middle town is 180 D; and the inner dimensions of the castle are 60 × 48 D. The sum of the width and length of the lower town comes to 729 which is astronomically significant since it is 27 × 27, and the width 324 equals the nakshatra year 27 × 12.

Continuity has been found between the grid and modular measures in the town planning of Harappa and historical India, including that of Kathmandu Valley (Pant and Funo, 2005). The measure of 19.2 meters is the unit in quarter-blocks of Kathmandu; this is nearly the same as the unit characteristic of the dimensions of Dholavira. It shows that the traditional architects and town planners have continued the use of the same units over this long time span.

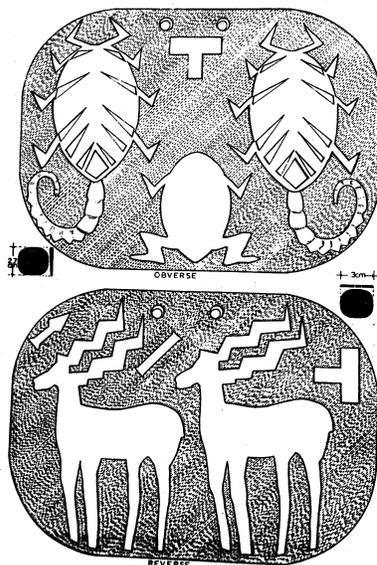

Figure 2. Astronomical seal from Rehman Dheri

*Rehman Dheri*
A 3rd millennium seal from Rehman Dheri, showing a pair of scorpions on one side and two antelopes on the other, that suggests knowledge of Vedic themes. It has been suggested that this seal represents the opposition of the Orion (Mrigashiras, or antelope head) and the Scorpio (Rohini of the southern hemisphere which is 14 nakshatras from the Rohini of the northern hemisphere) nakshatras. The arrow near the head of one of the antelopes could represent the decapitation of Orion. It is generally accepted that the myth of Prajapati being killed by Rudra represents the shifting of the beginning of the year away from Orion and it places the astronomical event in the fourth millennium BCE (Kak, 2000a).

## 3. Neolithic and Megalithic Sites

Interesting sites of archaeoastronomical interest include the Neolithic site of Burzahom from Kashmir in North India, and megalithic sites from Brahmagiri and Hanamsagar from Karnataka in South India.

*Burzahom, Kashmir*
This Neolithic site is located about 10 km northeast of Srinagar in the Kashmir Valley on a terrace of Late Pleistocene-Holocene deposits. Dated to around 3000 - 1500 BCE, its deep pit dwellings are associated with ground stone axes, bone tools, and gray burnished pottery. A stone slab of 48 cm × 27 cm, obtained from a phase dated to 2125 BCE shows two bright objects in the sky with a hunting scene in the foreground. These have been assumed to be a depiction of a double star system (Kameshwar Rao, 2005).

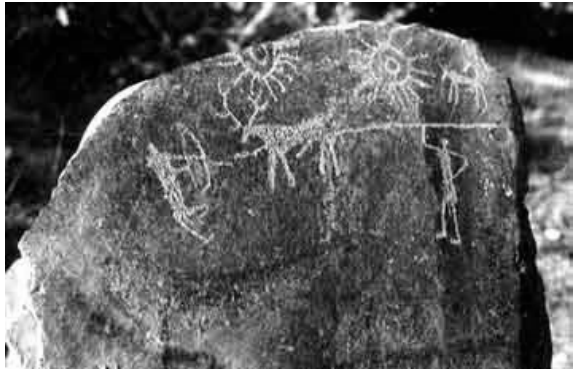
Figure 3. Burzahom sky scene

*Brahmagiri, Karnataka*
The megalithic stone circles of Brahmagiri in the Chitradurga district of Karnataka in South India, which have been dated to 900 BCE, show astronomical orientations. This site is close to Siddapur where two minor Aśokan rock edicts were found in 1891. Kameswara Rao (1993) has argued that site lines from the centre of a circle to an outer tangent of another circle point to the directions of the sunrise and full moon rise at the time of the solar and lunar solstices and equinox.

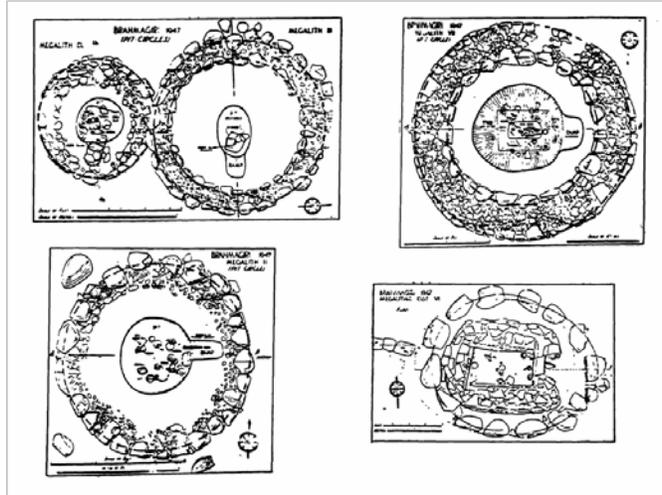
Figure 4. Megalithic stone circles of Brahmagiri

*Hanamsagar, Karnataka.*
Hanamsagar is a megalithic site with stone alignments pointing to cardinal directions. It is located on a flat area between hills about 6 km north of the Krishna river at latitude 16º 19′ 18″ and longitude 76º 27′ 10″. The stones, which are smooth granite, are arranged in a square of side that is about 600 meters with 50 rows and 50 column (for a total of 2,500 stones), with a separation between stones of about 12 m. The stones are between 1 to 2.5 m in height with a maximum diameter of 2 to 3 m. The lines are oriented in cardinal directions. There is a squarish central structure known as *chakri katti*.

It has been argued that the directions of summer and winter solstice can be fixed in relation to the outer and the inner squares. Kameswara Rao (2005) suggests that it could have been used for several other kind of astronomical observations such as use of shadows to tell the time of the day, the prediction of months, seasons and passage of the year.

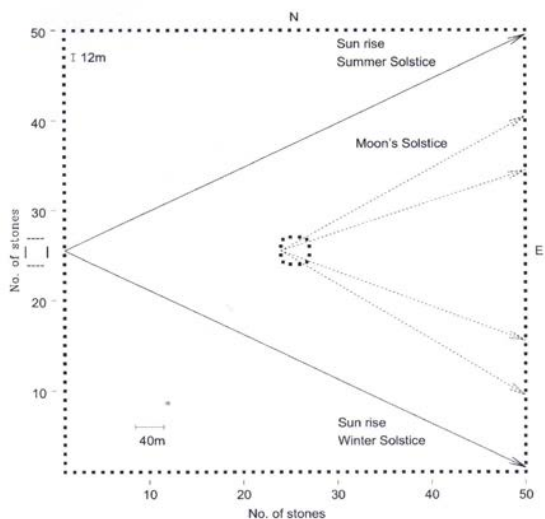
Figure 5. Alignments at Hanamsagar

## 4. The Plan of the Temple

The sacred ground for Vedic ritual is the precursor to the temple. The Vedic observances were connected with the circuits of the sun and the moon (Kak, 1993; Kak, 1995; Kak, 1996). The altar ritual was associated with the east-west axis and we can trace its origins to priests who maintained different day counts with respect to the solstices and the equinoxes. Specific days were marked with ritual observances that were done at different times of the day.

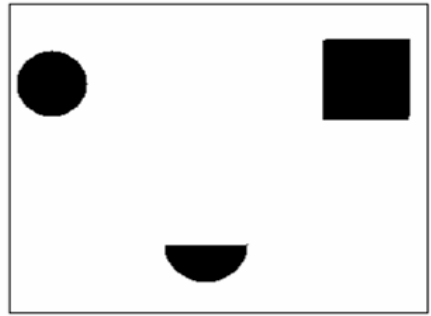

Figure 6. The three altars of the Vedic house: circular
(earth, body), half-moon (atmosphere, prāna), square (sky, consciousness)

In the ritual at home, the householder employed three altars that are circular (earth), half-moon (atmosphere), and square (sky), which are like the head, the heart, and the body of the Cosmic Man (*Purusha*). In the Agnichayana, the great ritual of the Vedic times that forms a major portion of the narrative of the *Yajurveda*, the atmosphere and the sky altars are built afresh in a great ceremony to the east. This ritual is based upon the Vedic division of the universe into three parts of earth, atmosphere, and sky that are assigned numbers 21, 78, and 261, respectively. The numerical mapping is maintained by placement of 21 pebbles around the earth altar, sets of 13 pebbles around each of 6 intermediate (13×6=78) altars, and 261 pebbles around the great new sky altar called the Uttara-vedi, which is built in the shape of a falcon; these numbers add up to 360, which is symbolic representation of the year. The proportions related to these three numbers, and others related to the motions of the planets, and angles related to the sightings of specific stars are reflected in the plans of the temples of the historical period (Kak, 2002b; Kak, 2006a; Kak, 2009; Kaulācara, 1966).

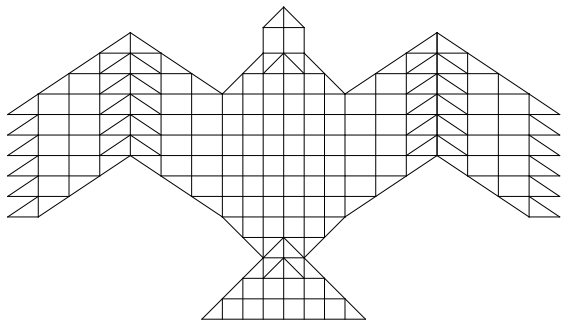

Figure 7. The falcon altar of the Agnichayana altar

The Agnichayana altar is the prototype of the temple and of the tradition of architecture (*Vāstu*). The altar is first built of 1,000 bricks in five layers (that symbolically represent the five divisions of the year, the five physical elements, as well as five senses) to specific designs. The altar is constructed in a sequence of 95 years, whose details are matched to the reconciliation of the lunar and solar years by means of intercalary months.

In the ritual ground related to the Agnichayana ceremony, the Uttara-vedi is 54 units from the entrance in the west and the perimeter of the ritual ground is 180 units (Kak, 2005a). These proportions characterize many later temples.

*The Temple Complex at Khajuraho*

The town of Khajuraho extends between 79° 54′ 30″ to 79° 56′ 30″ East and 24° 50′ 20″ to 24° 51′ 40″ North, in Chhatarpur district, in Madhya Pradesh. The temples of Khajuraho were built in 9th -12th century CE by the Chandela kings. Originally there were 84 temples, of which 23 have survived. Of the surviving temples, 6 are associated with Shiva, 8 with Vishnu, and 5 with the goddess (Singh, 2009b).

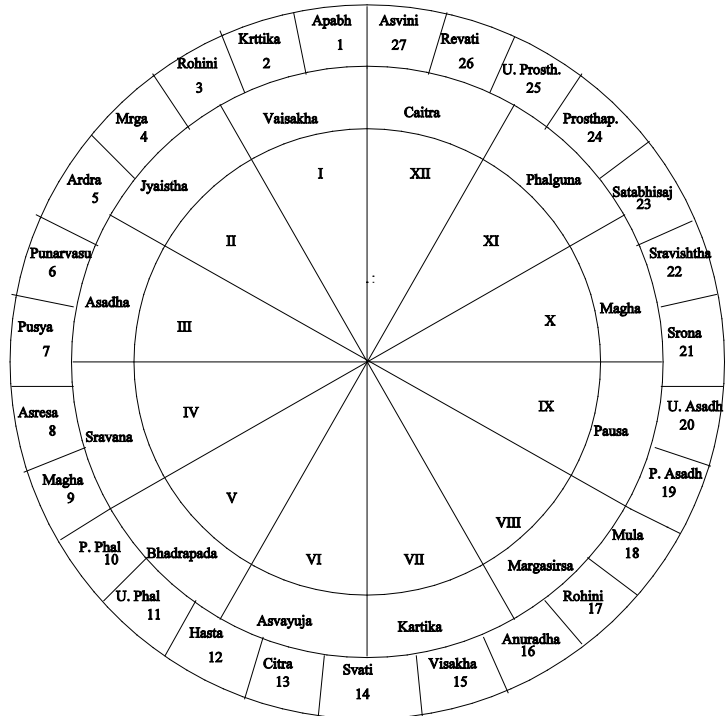

Figure 8. Mapping of the nakshatras to the solar months

At the eastern edge of the temple complex are the Dantla hills, with a peak of 390 m at which is located a shrine to Shiva, which is a reference point for the temple entrances. All the temples excepting the Chaturbhuja face the east. The southeastern edge has the Lavanya hill that is separated from the Dantla hills by the eastward flowing river Khudar. At the foothills of the Lavanya hill at a height of 244m is the shrine of goddess Durga as Mahishasurmardini.

The shrines to Shiva and Durga on the Dantla and Lavanya hills span the polarities of spirit (*Purusha*) and matter (*Prakriti*), which are bridged by the river between the hills. The temples of Khajuraho are popular pilgrimage centers during two spring festivals: Shivaratri that falls on the new moon of Phalguna (February/March), and Holi, which falls on the full moon of Chaitra (March/April).

The Lakshmana temple, one of the oldest of the complex, is considered the *axis mundi* of the site. It was built by the king Yashovarman (925-950) as symbol of the Chandela victory over the Pratiharas and a record of supremacy of their power. This temple is oriented to the sunrise on Holi.

The groups of temples form three overlapping mandalas, with centers at the Lakshmana (Vishnu), the Javeri (Shiva), and the Duladeva (Shiva) temples. Their deviation from true cardinality is believed to be due to the direction of sunrise on the day of consecration (Singh, 2009).

The temple, as a representation of the cosmos and its order, balances the *asuras* (demons) and the *devas* (gods), as well as inheres in itself other polarities of existence. In the Lakshmana Temple, Vishnu is depicted in a composite form with the usual calm face bracketed by the faces of lion and boar. The conception of the sanctum is as a mandala (Desai, 2004).

The planetary deities, the *grahas*, encircle the temple in the following arrangement:

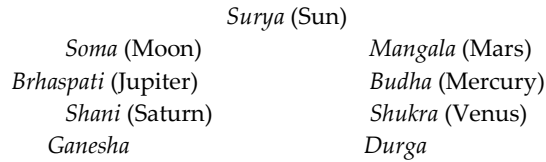

Ganesha and Durga are the deities of the ascending and the descending nodes of the moon, respectively. The temple is envisioned like Mount Meru, the axis of the universe, and the planets move around it.

## 5. The Udayagiri Observatory

Udayagiri ("hill of [sun]-rise"] is one of the principal ancient astronomical observatories of India. It is located at 23°31' N latitude on the Tropic of Cancer in Madhya Pradesh, about 50 kilometers from Bhopal, near Vidisha, Besnagar and Sanchi. An ancient site that goes back to at least the second century BCE, it was substantially enlarged during the reign of the Gupta Emperor Chandragupta II Vikramaditya (r. 375-414). This site is associated with 20 cave temples that have been cut into rock; nineteen of these temples are from the period of Chandragupta's reign (Dass and Willis, 2002).

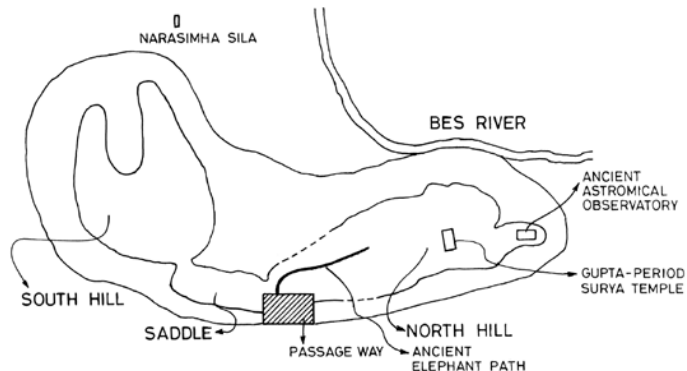
Figure 9. Udayagiri layout (Balasubramaniam, 2008)

It appears that the ancient name of Udayagiri was Vishnupadagiri, or the "hill of the footprint of Vishnu," and the name Udayagiri is after the Paramara ruler Udayaditya (c. 1070-93). The hill is shaped like a foot. A saddle connects the northern and southern hills, and a passageway is located at the place where the northern hill meets the saddle. The Gupta period additions and embellishments at Udayagiri were concentrated around this passage. Most of the cave temples are located around the passageway.

On the summer solstice day, there was an alignment of the sun's movement with the passageway. The day mentioned in the dated Chandragupta II Vikramaditya period inscription in cave 6 has been calculated to be very close to the summer solstice of the year 402 CE. On this day, the shadow of the Iron Pillar of Delhi, which was originally located at the entrance of the passageway, fell in the direction of the reclining Vishnu panel (Balasubramaniam, 2008).

On the northern hilltop, there exists a flat platform commanding a majestic view of the sky. Several astronomical marks have been identified at this platform, indicating that this was the site of the ancient astronomical observatory.

## 6. Medieval Pilgrimage Complexes

Medieval pilgrimage centers fulfilled many functions including that of trade and business. They were important to the *jyotishi* (astrologer) who would make and read the pilgrims' horoscope. The better astrologers were also interested in astronomy and this knowledge was essential for the alignment of temples and palaces.

Every region of India has important pilgrimage centers, some of which are regional and others pan-Indic. The most famous of the pan-Indic centers are associated with Shiva (Varanasi), Krishna (Mathura, Dwarka), Rama (Ayodhya), Vishnu (Tirupati), and the 12-yearly rotation of the Kumbha Mela at Prayag, Haridwar, Ujjain, and Nashik. For pilgrimage centers such as Chitrakut, Gaya, Madurai, Varanasi, Vindhyachal, and Khajuraho, the question of alignments of temples to cardinal directions or to direction of the sun on major festivals has been studied by scholars (Singh, 2009b). Here we will consider the sun temples of Varanasi (Malville, 1985; Singh, 2009a and 2009b).

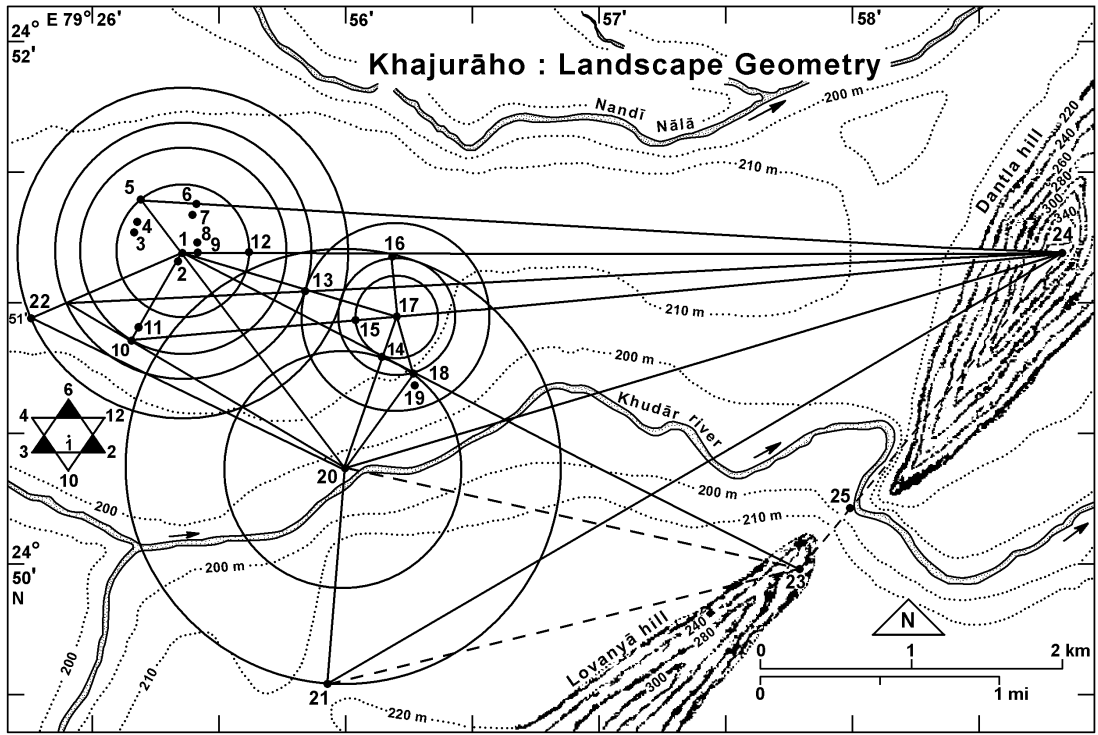

Figure 10. Khajuraho: Landscape Geometry and Topography (Singh, 2009b).

*The Sun Temples of Varanasi*
Varanasi is an ancient city dating from the beginning of the first millennium BCE, whose Vedic name is Kashi (Sanskrit for "radiance"), a name that continues to be used together with Banaras. Of its many temples, the most important is Kashi Vishvanath Temple, or "Golden Temple," dedicated to Lord

Shiva, the presiding deity of the city. Because of repeated destruction by the sultans and later by Aurangzeb, the current Vishvanath is a relatively modern building. It was built in 1777 by Maharani Ahilyabai of Indore, and its *shikhara* (spire) and ceilings were plated with of gold in 1839, which was a gift from Maharaja Ranjit Singh (Singh, 2009a and 2009b).

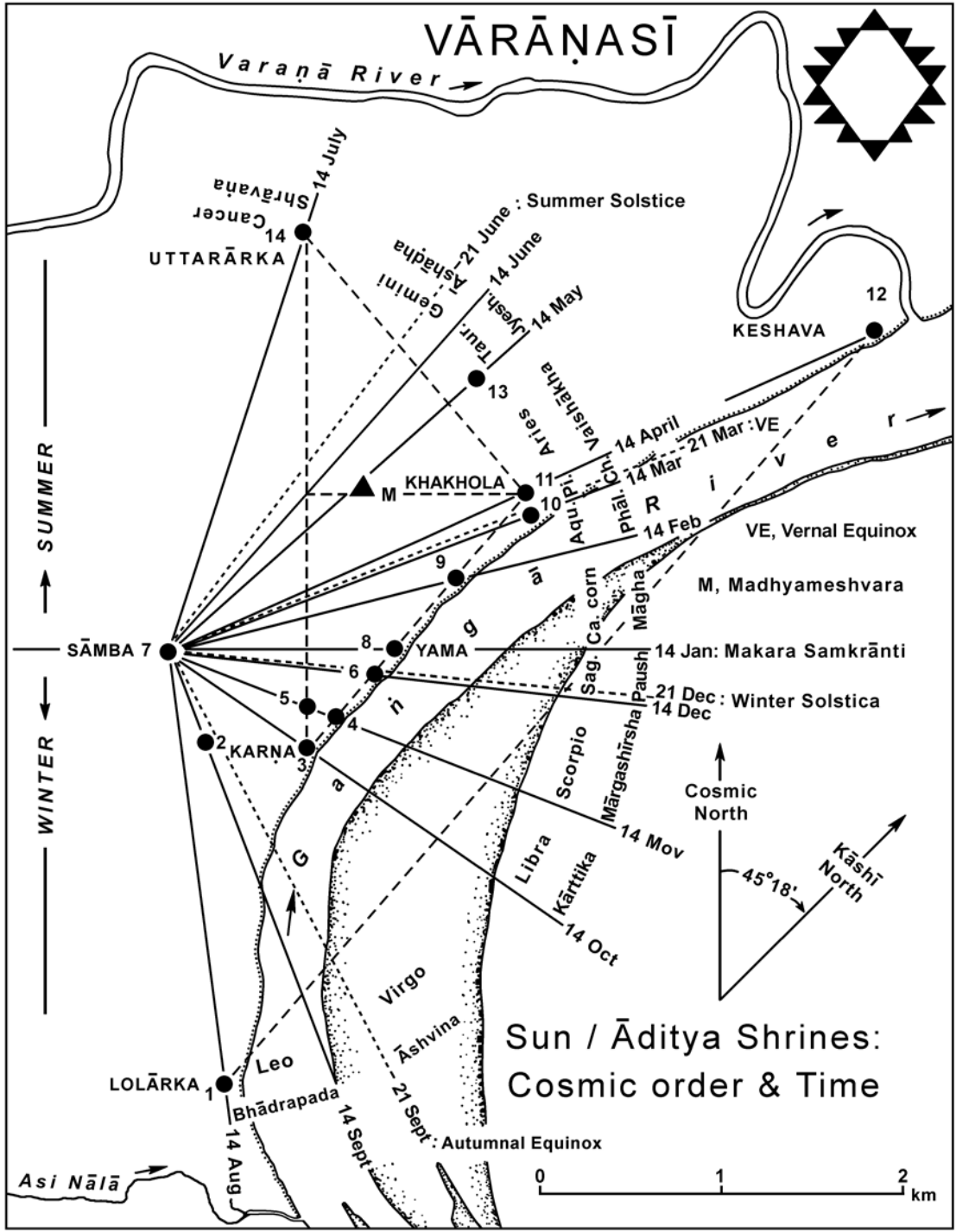

Figure 11. Sun Shrines: Cosmic Order and Cyclic orientation of Time (Singh, 2009a).

*Archaeoastronomy in India – Subhash Kak* 12

Shiva represents both the axis of the universe as well as that of one's inner being. One of the great festivals celebrated in Varanasi is Shivaratri which is celebrated on the 13th day of the dark fortnight of the Phalguna month (February-March). On that day you can see the sun rise in the east with the new moon just above it, which is represented iconographically by Shiva (as the sun) wearing the moon on his head.

There are several pilgrimage circuits in Varanasi for circumambulating the city. The Panchakroshi circuit has 108 shrines on it, and the four inner circuits have a total of 324 shrines. It is also known for the circuit of the Aditya shrines. The Adityas are the 7 or 8 celestial gods, although their number is counted to 12 in later books. In Puranic India, they are taken to be the deities of the twelve solar months. The Aditya temples were also razed during the centuries of Muslim rule, but have been re-established at the same sites and are now part of the active ritualscapes (Singh, 2009a).

Several Aditya shrines have been located with the aid of descriptions in the *Kashi Khanda* and pilgrimage guides (Singh and Malville, 1995; Singh, 2009a and 2009b). Six of these lie along one sides of an isosceles triangle with a base of 2.5km. The triangle surrounds the former temple of Madhyameshavara, which was the original center of Kashi. Pilgrims walking along the triangle are symbolically circumambulating the cosmos.

## 7. Sacred Cities

There are numerous sacred cities in the Indian sub-continent that were either built to an archetypal master plan or grew organically by virtue of being connected to a specific celestial deity. Some of the important sacred cities are:

1. Varanasi
2. Vijayanagara
3. Ayodhya
4. Mathura
5. Bhaktapur
6. Tirupati
7. Kanchipuram
8. Dwarka
9. Ujjain

Robert Levy viewed the Indian sacred city as a structured "mesocosm", situated between the microcosm of the individual and the macrocosm of the culturally conceived larger universe (Levy, 1991). Such a city is constructed of spatial connected *mandalas*, each of which is sustained by its own culture and performance. The movements of the festival year and rites of passage constitute a "civic dance", which defines the experience of its citizens.

The life-cycle passages and festivals dedicated to the gods affirm the householders' moral compass, identities and relationships. But there also exist other deities, represented generally by goddesses, who point to the forces of nature outside of moral order. These are brought into the larger order through tantric invocations and amoral propitiatory offerings. Performances invoking the goddess are the responsibility of the king and the merchants.

*Sacrality and Royal Power at Vijayanagara*
The city of Vijayanagara (also known as Hampi) was founded in the 14th century and sacked in 1565. The best known kings associated with Vijayanagara are Harihara I and II and Bukka Raya I (ca. 1336-1404), and Krishnadevaraya and his half-brother Achyutadevaraya (1509-42). From the mid-14th century to 1565, the city was the capital of the Vijayanagara Empire. According to the Persian ambassador Abdur Razaaq (1442 CE): "The City of Vijayanagara is such that the pupil of the eye has never seen such a place like it, and the ear of intelligence has never been informed that there existed anything to equal it in the world."

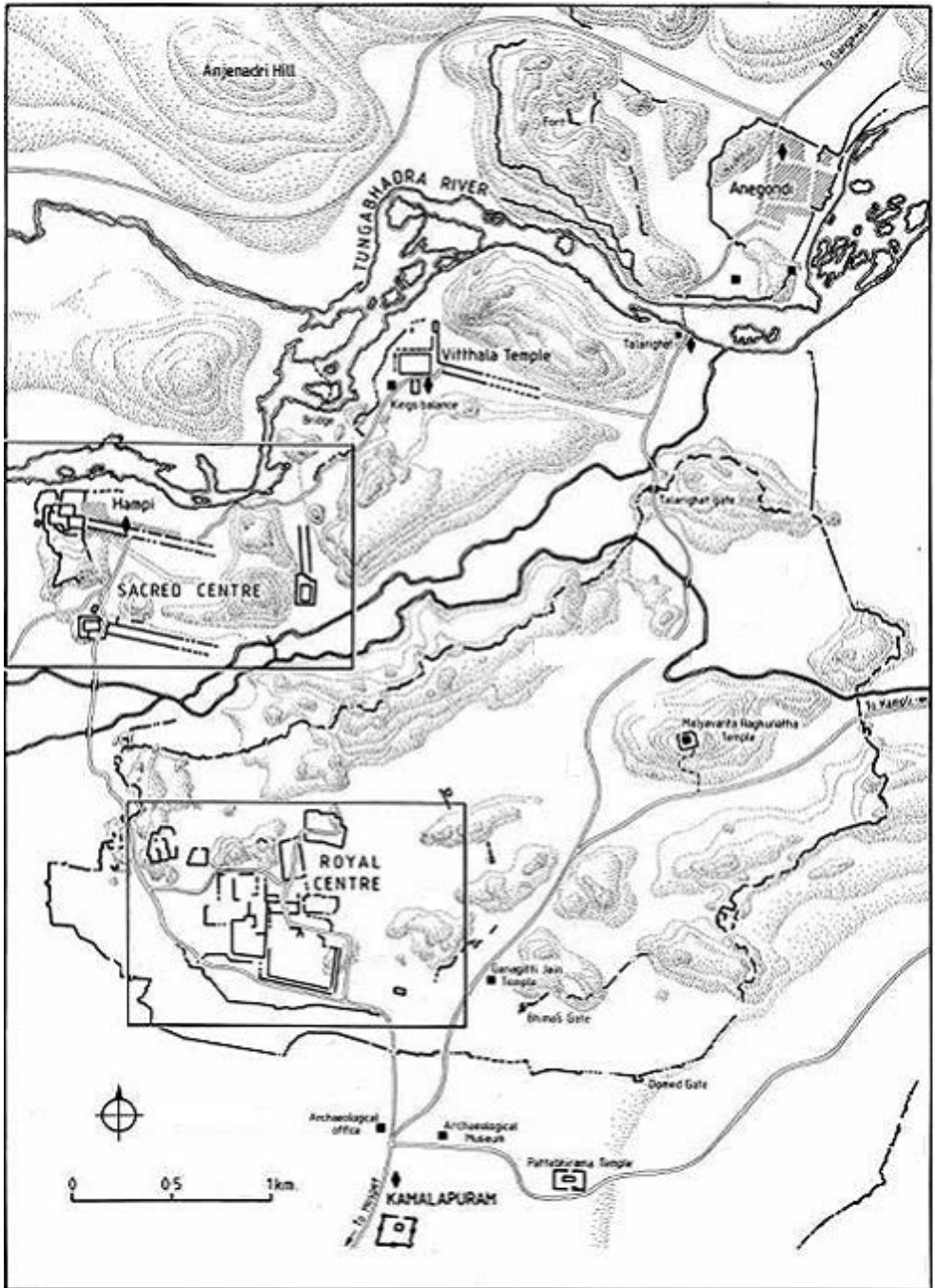

Figure 12. Vijayanagara City

Hampi had for centuries been an important pilgrimage city due to its mythic association with river Goddess Pampā and her consort Virupaksha, or Pampāpati. An inscription dated 1163 CE records a *mahādāna*, a religious offering in the presence of Lord Virupaksha of Hampi by the Kalachuri King Bijjala. The region was part of the kingdom of Kampiladeva until 1326 when the armies of Mohammed Bin Tughlaq defeated the king and imprisoned the two sons of Sangama, Hukka and Bukka. Some years later the Sultan sent the two as governors of the province. In 1336 they broke free from Tughlaq allegiance and established the Sangama dynasty with its capital at Vijayanagara.

The destruction of Vijayanagara in 1565 was captured vividly in the account of Robert Sewell (1900): "They slaughtered the people without mercy; broke down the temples and palaces; and wreaked such savage vengeance on the abode of the kings that, with the exception of a few great stone built temples and walls, nothing now remains but a heap of ruins to mark the spot where once the stately buildings stood… They lit huge fires in the magnificently decorated buildings forming the temple of Vitthalaswami near the river, and smashed its exquisite stone sculptures. With fire and sword, crowbars and axes they carried on day after day their work of destruction. Never perhaps in the history of the world has such havoc been wrought so suddenly on so splendid a city; teeming with a wealthy and industrious population in the full plenitude of prosperity one day and on the next seized, pillaged and reduced to ruins amid scenes of savage massacre and horrors beggaring description."

Hampi has a strong association with the Ramayana and the names of many sites in the area bear names mentioned in the epic. These include Rishimukha, Malyavanta hill and Matanga hill along with a cave where Sugriva is said to have kept the jewels of Sita. The site of Anegundi is associated with the kingdom of Angad, son of Vali. The Anjaneya Parvata, a hill to the west of Anegundi, is the fabled birthplace of Hanuman.

Hampi is also linked with the river goddess Pampā and the legend of her marriage to Lord Virupaksha or Shiva. Each year, in the month of Chaitra (March-April), this marriage is re-enacted, with the priests of Virupaksha temple devoutly performing every ritual from Phalapūjā (betrothal) to Kalyānotsava (marriage) in the temple.

The Sacred Center of the city lies south of the Tungabhadra River, and it is dominated by four large complexes of the Virupaksha, Krishna, Tiruvengalanatha (Achyutaraya) and Vitthala temples. The major temples are either close to cardinality, departing by an average of 10', or are oriented to major features of the sacred landscape.

Further south of the Sacred Center is the Royal Center, which is divided into the public and private realms. The division is achieved by a north-south axis, which passes almost precisely between the king's 100-column audience hall in the east and the queen's large palace in the west. The Ramachandra temple pierces the axis by connecting the private and the public domains. In the homology of the king and the deity, the king is able to inhere in him the royalty and divinity of Rama.

The Virabhadra temple is on the summit of Matanga hill, which is the center of the *vāstu-mandala* and the symbolic source of protection that extended outward from it along radial lines. As viewed from a point midway between the audience hall and the queen's palace, the *shikhara* of the Virabhadra lies only 4 minutes of arc (4') from true north. The ceremonial gateway in the corridor west of Ramachandra temple joined with the summit of Matanga hill departs from true north by 0.6 minutes of arc (0.6') (Malville, 2000).

The orientations of the major axes of the small temples, shrines, and palaces of the urban core are in marked contrast to those. The smaller structures are rotated away from cardinality for the four directions by $17^{o}$, suggesting that they were influenced by the position of the rising sun on the morning when it crosses the zenith.

The bazaar streets of the Virupaksha, Vitthala and Krishna temples are set between 13 and 15 degrees south of east. Malville (2000) speculates that there may be some link between these orientations and the rising point of the star Sirius.

8. **Conclusions**

Interest in archaeoastronomy and art, as connected to temples and ancient monuments, has increased in India as the country's prosperity has increased. This increase is also owing to the major archaeological discoveries that have been made in the past few decades and the importance of temple tourism.

The principal authority over significant sites is the Indian Archaeological Survey of India (ASI) and its sister institutions that function at the state level as Departments of Archaeology and Museums. In 1976, the Indian Government initiated projects to excavate three great medieval cities: Fatehpur Sikri in Uttar Pradesh, Champaner in Gujarat, and Vijayanagara in Karnataka, which are UNESCO World Heritage sites. The wealth of discoveries made in these cities is strengthening the movement to expose and preserve other sites in the country. The efforts at excavation, conservation, and research can only be expected to increase. In particular, greater attention will be given to the archaeoastronomical aspects of the monuments.

**Acknowledgements.** I am thankful to R. Balasubramaniam, Michel Danino, McKim Malville, and Rana P.B. Singh for their advice. The essay is dedicated to the memory of R. Balasubramaniam who passed away in December 2009.

This article is the unabridged version of an essay written for the International Council on Monuments and Sites (ICOMOS). An abridged version appeared as a chapter titled "India" on pages 101-110 in the book "Heritage Sites of Astronomy and Archaeoastronomy in the context of the UNESCO World Heritage Convention: A Thematic Study" edited by Clive Ruggles and Michel Cotte that was published jointly by the International Council on Monuments and Sites (ICOMOS) and the International Astronomical Union (IAU), Paris, in 2010.

## Select Bibliography